\documentclass[twocolumn,aps,prb,showpacs]{revtex4}

\usepackage{epsfig}

\begin{document}

%\draft

\title{Density of state and non-magnetic impurity effects in electron-doped
  cuprates}
\author{Bin Liu$^1$ and Ying Liang$^2$ }

\affiliation{$^1$ Max-Planck-Institut f\"ur Physik komplexer
Systeme, D-01187 Dresden, Germany \\
$^2$ Department of Physics, Beijing Normal University, Beijing 100875, China}

%\date{\today}
\begin{abstract}

we analyze the density of state (DOS) and a non-magnetic impurity
effect in electron-doped cuprates starting from two different
scenarios: the $d_{x^{2}-y^{2}}$-wave superconductivity coexisting
with antiferromagnetic spin density wave (SDW) order versus
$d_{x^{2}-y^{2}}$-wave superconductivity with a higher harmonic. We
find that in both cases the local density of state (LDOS) exhibits
two impurity-induced resonance states at low energies. We also find
that for the intermediate value of the SDW gap, the DOS looks
similar to that obtained from the scenario of the
$d_{x^{2}-y^{2}}$-wave gap with a higher harmonic, suggesting the
presence of a non-monotonic $d_{x^{2}-y^{2}}$-wave gap. However, if
the SDW gap is sufficiently large the DOS looks more conventional
s-wave like. This obvious difference from the DOS resulted from the
$d_{x^{2}-y^{2}}$-wave gap with a higher harmonic model, could
differentiate the two above scenarios and is needed to be proved in
the further doping dependence of tunneling spectrum measurement.

\end{abstract}
\pacs{74.72.Jt, 74.20.Mn, 74.25.Ha, 74.25.Jb}

\maketitle

%\narrowtext
\section{ Introduction}

The electron-doped cuprate high-temperature superconductors have
recently attracted a considerable attention due to their
considerable asymmetry from hole-doped ones. For instance, it has
been known that in contrast to the hole-doped cuprates, the
electron-doped systems show (i) relatively low superconducting
transition temperature, (ii) narrow superconducting (SC) region in
the phase diagram, and (iii) the robust antiferromagnetic (AF)
order\cite{millis}.

The phase sensitive scanning SQUID measurements\cite{tsuei}, nuclear magnetic
resonance study\cite{zheng}, and also
ARPES experiments\cite{matsui,arm} have provided strong evidences
that the electron-doped superconductors are the
$d_{x^{2}-y^{2}}$-wave superconductors. This conclusion seems to be
natural, since from the theoretical point of view, the same pairing
mechanism is expected both for hole- and electron-doped materials,
although the functional form of the $d_{x^{2}-y^{2}}$-wave gap in
electron-doped materials is more subtle issue. The high-resolution
ARPES data on the leading-edge gap in
$Pr_{0.89}LaCe_{0.11}CuO_{4}$\cite{matsui}, Raman experiments in
$NCCO$\cite{blumberg}, and doping dependence of tunneling study in
$Pr_{2-x}Ce_{x}CuO_{4-\delta}$\cite{dagan} show a non-monotonic
$d_{x^{2}-y^{2}}$-wave gap with a maximum value in between nodal and
antinodal points on the Fermi surface (FS), and meanwhile the
measurements of optical conductivity $\sigma_{1}(\omega)$ in
$Pr_{1.85}Ce_{0.15}CuO_{4}$\cite{homes} were also interpreted as an
indirect evidence of a non-monotonic $d_{x^{2}-y^{2}}$-wave gap.

The theoretical explanations for the origin of the non-monotonic gap
behavior are classified into two scenarios so far. One is the
coexisting AF with SC order
scenario\cite{yoshi,yuan}. Although the SC gap itself is assumed to be
monotonic $d_{x^{2}-y^{2}}$-wave, when the AF order is introduced, the
resulting quasiparticle excitation can be gapped by both orders and behaves to
be non-monotonic $d_{x^{2}-y^{2}}$-wave gap. The other one is to
extend the SC gap out of the simplest $d_{x^{2}-y^{2}}$-wave via the
inclusion of a higher harmonic term\cite{manske,liu2}. From theory
perspective, the non-monotonic $d_{x^{2}-y^{2}}$-wave gap appears naturally under the assumption that the $d_{x^{2}-y^{2}}$-wave pairing is caused by the
interaction with the continuum of overdamped AF spin
fluctuations. Spin-mediated interaction is attractive in the $d_{x^{2}-y^{2}}$-wave channel and yields a gap which is maximal near the
hot spots - the points along the FS, separated by AF moment $Q_{AF}$. In
optimally doped $NCCO$ and $PCCO$, hot spots are located close to Brillouin
zone diagonals, and one should generally expect the $d_{x^{2}-y^{2}}$-wave gap
to be non-monotonic. In this case, the non-monotonic gap behavior
is an intrinsic property in the SC state regardless of the presence of the AF order.

In fact, impurity effect has always been used as one of the most
important and effective tools to distinguish pairing symmetry in the
conventional and unconventional superconductors. It has been known
that the Yu-Shiba-Rusinov state in the conventional BCS
superconductor (s-wave) was located at the gap edge\cite{yu}.
However in the $d_{x^{2}-y^{2}}$-wave superconductor an
impurity-induced bound state appear near the Fermi
energy\cite{wang}. The origin of this difference has been explained
as different phase structure of two kinds of SC order:
$d_{x^{2}-y^{2}}$-wave pairing symmetry with the line nodal gap
while s-wave symmetry with nodeless gap. Therefore, we in this paper
analyze the DOS and a non-magnetic impurity effect starting from two
different scenarios: the $d_{x^{2}-y^{2}}$-wave superconductivity
coexisting with AF SDW wave order versus $d_{x^{2}-y^{2}}$-wave
superconductivity with a higher harmonic. We find that in both cases
the LDOS exhibits two impurity-induced resonance states at low
energies. We also find that for the intermediate value of the SDW
gap, the DOS looks similar to that obtained from the scenario of the
$d_{x^{2}-y^{2}}$-wave gap with a higher harmonic, suggesting the
presence of a non-monotonic $d_{x^{2}-y^{2}}$-wave gap. However, if
the SDW gap is sufficiently large the DOS looks more conventional
s-wave like instead of the non-monotonic $d_{x^{2}-y^{2}}$-wave gap
behavior. This obvious difference from the DOS resulted from the
$d_{x^{2}-y^{2}}$-wave gap with a higher harmonic model even in the
underdoped regimes, could differentiate the two above scenarios. Our
results strongly suggest the further doping dependence of tunneling
spectrum measurement, especially in the underdoped regimes, is
needed to be carried out so as to shed light on the physical origin
of the unusual non-monotonic gap in the electron-doped cuprates.

%%%%%%%%%%%%%%%%%%%%%%%%%%%%%%%%%%%%%%%%%%%%%%%%%%
\section{the Model and T-matrix formulation}
%%%%%%%%%%%%%%%%%%%%%%%%%%%%%%%%%%%%%%%%%%%%%%%%%%

We start from a phenomenological superconducting Hamiltonian on a
square lattice at the mean field level,
\begin{eqnarray}
H&=&H_{SC}+H_{SDW}, \nonumber\\
H_{SC}&=&\sum_{{\bf k}\sigma}[(\varepsilon_{\bf k}-\mu)c_{{\bf
k}\sigma}^{\dagger}c_{{\bf k}\sigma} \nonumber\\&+&\Delta_{\bf k}(c_{{\bf
k}\uparrow}^{\dagger}c_{-{\bf k}\downarrow}^{\dagger}+c_{-{\bf
k}\downarrow}c_{{\bf k}\uparrow})], \nonumber\\
H_{SDW}&=&-\sum_{{\bf k}\sigma}M\sigma(c_{{\bf k}\sigma}^{\dagger}c_{{\bf
k+Q}\sigma}+h.c.),
\end{eqnarray}
where $c_{{\bf k}\sigma}^{\dagger}$ ($c_{{\bf k}\sigma}$) is the
fermion creation (destruction) operator and $\mu$ is chemical
potential. The SC order parameter has the usual form $\Delta_{\bf
k}=\Delta(\cos(k_{x})-\cos(k_{y}))/2$ and the AF SDW order parameter
is $M$. We take the normal state tight binding dispersion
$\varepsilon_{\bf k}$, which is written as
\begin{eqnarray}
\varepsilon_{\bf
k}&=&-2t(\cos(k_{x})+\cos(k_{y}))-4t_{1}\cos(k_{x})\cos(k_{y})\nonumber\\
&-&2t_{2}(\cos(2k_{x})+\cos(2k_{y}))
 \nonumber\\
&-&4t_{3}(\cos(2k_{x})\cos(k_{y})+\cos(k_{x})\cos(2k_{y}))\nonumber\\
&-&4t_{4}\cos(2k_{x})\cos(2k_{y})
\end{eqnarray}
%\begin{eqnarray}
%\varepsilon_{\bf
%k}&=&-2t(\cos(k_{x})+\cos(k_{y}))-4t_{1}\cos(k_{x})\cos(k_{y})
% \nonumber\\
%&-&2t_{2}(\cos(2k_{x})+\cos(2k_{y}))
%\end{eqnarray}
with $t=120$ meV, $t_{1}=-60$ meV, $t_{2}=34$ meV, $t_{3}=7$ meV,
$t_{4}=20$ meV, and $\mu=-82$ meV at 0.11 doping\cite{bansil1}
reproducing the underlying FS as inferred from recent ARPES
experiment\cite{matsui}. Note that here the wave vector ${\bf k}$ is
restricted to the magnetic Brillouin zone (MBZ).

It is convenient to introduce a $4\times4$ matrix formulation to treat the
coexisting SDW and SC phase. After taking a four-component Nambu spinor
$\Psi_{{\bf k}}=(c_{{\bf k}\uparrow},c_{{\bf k+Q}\uparrow},c_{-{\bf
k}\downarrow}^{\dagger},c_{-{\bf k-Q}\downarrow}^{\dagger})^{\top}
$
with ${\bf Q}=(\pi,\pi)$ being the nesting vector, we rewrite the
Hamiltonian as
\begin{eqnarray}
H=\sum_{{\bf k}}\Psi^{+}_{{\bf k}}((\varepsilon_{\bf k}-\mu)\tau_{3}\rho_{0}+M\tau_{1}\rho_{0}+\Delta_{\bf
k}\tau_{3}\rho_{1})\Psi_{{\bf k}},
\end{eqnarray}
where $\tau_{3}\rho_{1}= \left (\matrix{0 &\tau_{3}\cr \tau_{3}
&0\cr}\right)$.
Then the single-particle matrix Green's function is determined as
$G_{0}^{-1}(k,i\omega_{n})=i\omega_{n}-(\varepsilon_{\bf k}-\mu)\tau_{3}\rho_{0}-M\tau_{1}\rho_{0}-\Delta_{\bf
k}\tau_{3}\rho_{1}$.

The scattering of quasiparticles from the impurity is described by a  T-matrix\cite{wang,fisher,wang1},
$T(i\omega_{n})$, which is local and independent of wave vectors. Thus, we define the $2\times2$ Green's function in the presence of a single impurity as
\begin{eqnarray}
G(i,j;i\omega_{n})&=&\zeta_{0}(i-j;i\omega_{n}) \nonumber\\
&+&\zeta_{0}(i,i\omega_{n})T(i\omega_{n})\zeta_{0}(j,i\omega_{n}),
\end{eqnarray}
where
\begin{eqnarray}
\zeta_{0}(i,j;i\omega_{n})&=&\frac{1}{N}\sum_{\bf k}e^{i\bf k\cdot\bf
R_{ij}}\nonumber\\
&\times& \left (\matrix{G^{1}_{0}(k,i\omega_{n})
&G^{2}_{0}(k,i\omega_{n})\cr G^{3}_{0}(k,i\omega_{n})
&G^{4}_{0}(k,i\omega_{n})\cr}\right),
\end{eqnarray}
with
\begin{eqnarray}
G^{{\bf I}}_{0}(k,i\omega_{n})&=&e^{-i\bf Q\cdot\bf
R_{j}}[G_{0}]_{{\bf I}}^{12}(k,i\omega_{n})+e^{i\bf Q\cdot\bf
R_{i}}[G_{0}]_{{\bf I}}^{21}(k,i\omega_{n})\nonumber\\&+&e^{i\bf
Q\cdot\bf R_{ij}}[G_{0}]_{{\bf I}}^{22}(k,i\omega_{n})+[G_{0}]_{{\bf
I}}^{11}(k,i\omega_{n}).
\end{eqnarray}
Here  ${\bf I}=1,2,3,4$ denotes the left-top, right-top, left-bottom
and right-bottom $2\times2$ block element of $G_{0}(k,i\omega_{n})$,
and $\bf R_{i}$ is lattice vector and $\bf R_{ij}=\bf R_{i}-\bf
R_{j}$. The T-matrix can be written by
\begin{eqnarray}
T(i\omega_{n})=\frac{U_{0}\rho_{3}}{1-U_{0}\rho_{3}\zeta_{0}(0,0;i\omega_{n})}.
\end{eqnarray}
For the d-wave pairing symmetry, one can find that the local Green's
function $\zeta_{0}(i,i;i\omega_{n})$ is diagonal. As a result, the
diagonal T-matrix
\begin{eqnarray}
T_{11,22}(i\omega_{n})=\frac{\pm U_{0}}{1-U_{0}[\zeta_{0}(0,0;\pm
i\omega_{n})]_{11}}
\end{eqnarray}
where the upper (lower) sign denotes $T_{11}$ ($T_{22}$), will give
rise to a particle- ($\omega_{res}<0$) and hole-like
($\omega_{res}>0$) resonance state. These resonance states generate the sharp
peaks in the LDOS only in the unitary limit
($\mid\omega_{res}\mid/\Delta\leq1$) where
$1=U_{0}Re[\zeta_{0}(0,0;\pm \omega_{res})]_{11}$.
Finally, these above equations allow a complete solution of the problem as
long as the order-parameter relaxation can be neglected.

%\begin{eqnarray}
%\zeta_{0}(i,j;i\omega_{n})=\frac{1}{N}\sum_{\bf k}e^{i\bf k\cdot\bf
%R_{ij}}\left (\matrix{\zeta^{11}_{0}(k,i\omega_{n})
%&\zeta^{12}_{0}(k,i\omega_{n})\cr \zeta^{21}_{0}(k,i\omega_{n})
%&\zeta^{22}_{0}(k,i\omega_{n})\cr}\right),
%\end{eqnarray}
%where
%\begin{mathletters}
%\begin{eqnarray}
%\zeta^{11}_{0}(k,i\omega_{n})&=&e^{-i\bf
%Q\cdot\bf R_{j}}G^{12}(k,i\omega_{n})+e^{i\bf Q\cdot\bf R_{i}}G^{21}(k,i\omega_{n})\nonumber\\
%&+&e^{i\bf Q\cdot\bf
%R_{ij}}G^{22}(k,i\omega_{n})+G^{11}(k,i\omega_{n});\\
%\zeta^{12}_{0}(k,i\omega_{n})&=&e^{-i\bf
%Q\cdot\bf R_{j}}G^{14}(k,i\omega_{n})+e^{i\bf Q\cdot\bf R_{i}}G^{23}(k,i\omega_{n}) \nonumber\\
%&+&e^{i\bf Q\cdot\bf
%R_{ij}}G^{24}(k,i\omega_{n})+G^{13}(k,i\omega_{n});\\
%\zeta^{21}_{0}(k,i\omega_{n})&=&e^{-i\bf
%Q\cdot\bf R_{j}}G^{32}(k,i\omega_{n})+e^{i\bf Q\cdot\bf R_{i}}G^{41}(k,i\omega_{n}) \nonumber\\
%&+&e^{i\bf Q\cdot\bf
%R_{ij}}G^{42}(k,i\omega_{n})+G^{31}(k,i\omega_{n});\\
%\zeta^{22}_{0}(k,i\omega_{n})&=&e^{-i\bf
%Q\cdot\bf R_{j}}G^{34}(k,i\omega_{n})+e^{i\bf Q\cdot\bf R_{i}}G^{43}(k,i\omega_{n}) \nonumber\\
%&+&e^{i\bf Q\cdot\bf
%R_{ij}}G^{44}(k,i\omega_{n})+G^{33}(k,i\omega_{n}),
%\end{eqnarray}
%\end{mathletters}
%and $\bf R_{i}$ is lattice vector and $\bf R_{ij}=\bf R_{i}-\bf
%R_{j}$.

\section{Numerical Results and Discussions}

\subsection{$d_{x^{2}-y^{2}}$-wave coexisting with AF SDW order}

We firstly calculate the DOS
\begin{eqnarray}
\rho(\omega)=-\frac{1}{\pi}{\rm Im}\sum_{ik}G_{ii}(k,\omega)
\end{eqnarray}
plotted in Fig.1 for different SDW gap ($M$). In principle, SDW gap
($M$) and SC gap ($\Delta$) need to be solved self-consistently. For
simplicity, we in the following calculation choose $\Delta=0.005eV$
reproducing the ARPES estimate of SC gap in
$Pr_{0.89}LaCe_{0.11}CuO_{4}$\cite{matsui}, and different values of
SDW gap $M= 0.14, 0.12, 0.05$, and $0 (eV)$, where the maximal value
$M=0.14 (eV)$ is a self-consistent value from Ref.15 at doping
$x=0.11$, and the gradually decreasing  values with the independent
band dispersion (Eq.(2)) corresponds to the increasing
doping\cite{bansil1,bansil2,kang}. In this case, our SDW gap
dependence of DOS can be qualitatively equivalent to the doping
evolution of DOS.

In the coexisting AF SDW and SC state, there are
three features of interest. First, at low energies the DOS spectrum
remains d-wave like superconductor behavior for fairly large values
$M$. Second, with the increasing of $M$, besides the coherent peak located at
the SC gap, another two new ``coherent
peaks" appear symmetrically at positive and negative low energy, and the
positions of these peaks shift towards the Fermi energy, indicating the
presence of non-monotonic $d_{x^{2}-y^{2}}$-wave gap behavior. The third and last feature is that at
sufficiently large SDW gap M, the DOS shows the U-shaped behavior instead of
the non-monotonic $d_{x^{2}-y^{2}}$-wave gap behavior. This U-shaped DOS has
been observed in earlier point contact tunneling
spectrum and was explained as the character of s-wave
pairing\cite{alff,biswas,blumberg}, but in present case it is the result
of the coexisting SDW and SC orders, the difference between them
will be differentiated by considering a non-magnetic impurity effect.

\begin{figure}[ht]
\epsfxsize=3.5in\centerline{\epsffile{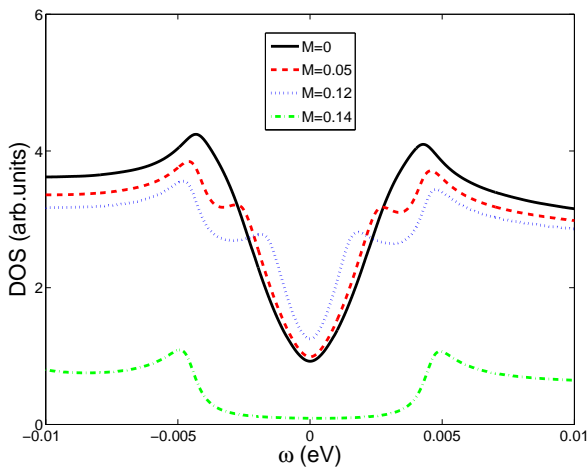}} \caption{ (color
online) The DOS for various SDW gap $M$. From bottom to top the SDW
gap are successively 0.14, 0.12, 0.05, and 0(eV).}
\end{figure}

These unusual SDW gap dependence of DOS can be understood as
follows. After diagonalizing the Hamiltonian (1), we obtain the
poles of $G(k,\omega)$, quasiparticle energy bands $\pm E^{\pm}_{\bf
k}$,
\begin{eqnarray}
E^{\pm}_{\bf k}=\sqrt{(\xi^{\pm}_{\bf k}-\mu)^{2}+\Delta^{2}_{\bf
k}}
\end{eqnarray}
with
\begin{eqnarray}
 \xi^{\pm}_{\bf k}=\frac{\varepsilon_{\bf k}+\varepsilon_{\bf
k+Q}}{2}\pm\sqrt{\frac{(\varepsilon_{\bf k}-\varepsilon_{\bf
k+Q})^{2}}{4}+M^{2}}
\end{eqnarray}
\begin{figure}[ht]
\epsfxsize=3.0in\centerline{\epsffile{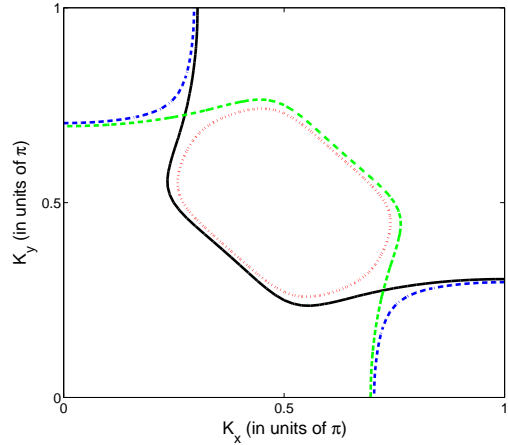}} \caption{ (color
online) Fermi surface for AF bands $(\xi^{\pm}_{\bf k}-\mu)$ with the
SDW gap $M=0.05$. The dotted-dashed curves are contributed by the
band $``+"$ and the dotted curve by band $``-"$. The zero-energy
contour of $\varepsilon_{\bf k}$ (solid curves) and
$\varepsilon_{\bf k+Q}$ (dashed curves) are shown for reference. }
\end{figure}
In the normal state, the Fermi surface for the corresponding AF
bands $(\xi^{\pm}_{\bf k}-\mu)$ has been plotted in Fig.2, where the
dotted-dashed curve is given by the band $``+"$ and the dotted curve by
band $``-"$. At the intermediate value of the SDW gap $M=0.05$, the
FS has been separated into three parts locating around $(\pi,0)$,
$(0,\pi)$ and $(\pi/2,\pi/2)$ respectively, which is quite agreement
with the ARPES experiment on $NCCO$ at the optimal
doping\cite{armitage} and the previous theoretical
calculations\cite{yuan,liu}. In this case, the DOS in the SC state
obviously shows two different coherent peaks: one at the low energy
corresponds to the SC gap maximum opened on the FS piece around
$(\pi/2,\pi/2)$, and the other at high energy corresponds to FS
pieces around  $(\pi,0)$, and $(0,\pi)$. Therefore the non-monotonic
$d_{x^{2}-y^{2}}$-wave behavior gap mentioned above can be
satisfactorily explained to the effective gap induced by coexisting
AF SDW and SC order. With the decreasing of the SDW, in the limit
case M=0 (solid curve in Fig.2), the two bands in the FS merge into
a single one again, therefore, the DOS shows a  V-shaped behavior,
similar to that of the hole-doped ones with a monotonic
gap\cite{arm,liu,blumberg,sne}. In contrast, when the SDW increases
to be sufficiently large (not shown here), the FS piece around
$(\pi/2,\pi/2)$ disappears. As a result, the DOS denoted by
dotted-dashed line in Fig.1 behaves to be U-shaped like. As expected above,
the SDW gap dependence of FS is qualitatively similar to the intriguing doping
evolution of FS\cite{armitage}.

\begin{figure}[ht]
\epsfxsize=3.5in\centerline{\epsffile{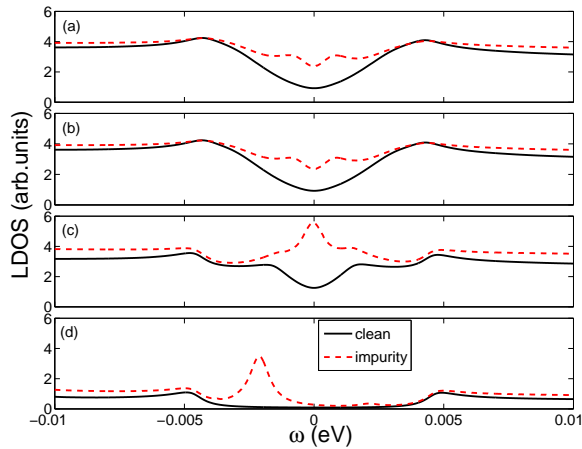}} \caption{(color
online)The LDOS for various SDW gap $M$ in the clean case and the
nearest neighbor of a non-magnetic impurity site. (a)$M=0$ eV,
(b)$M=0.05$ eV, (c)$M=0.12$ eV, (d)$M=0.14$ eV.}
\end{figure}

In the presence of a single non-magnetic impurity, the LDOS
which can be measured in the STM experiment
\begin{eqnarray}
N(r,\omega)&=&-\frac{1}{\pi}{\rm
    Im}G_{11}(r,r;\omega+i0^{\dagger}) \nonumber\\
&+&\frac{1}{\pi}{\rm Im}G_{22}(r,r;-\omega-i0^{\dagger})
\end{eqnarray}
with the subscripts 11 and 22 refer to the electron and hole part of
the diagonal Nambu Green's function, has been plotted in
Fig.3(a)-3(d) for different AF SDW gap, where black solid curve
denotes the clean case which is equivalent to DOS, and red dashed
curve denotes the nearest neighbor of the impurity site. In the
calculation, we take the same parameters as in Fig.1 and set the
scattering strength $U_{0}=5$eV. Note that our results remain
qualitatively unchanged when $U_{0}$ is varied. Since the DOS
remains $d_{x^{2}-y^{2}}$-wave like at low energies for fairly large
values $M$, the two impurity-induced resonance peaks in LDOS appear
at positive and negative energy and have different height due to the
particle-hole asymmetry, similar to the hole-doped cases\cite{wang}.
As $M$ increases, due to the presence of new ``coherence peaks" and
an inward shift in the position of them in DOS, two impurity-induced
resonance peaks correspondingly shift towards the lower energy and
eventually merge into one peak at Fermi energy. For the sufficiently
large $M$, although the DOS behaves to be U-shaped like, two
impurity-induced resonance peaks still exist, which is different
from the pure s-wave superconductors where no resonance state within
the gap was induced by a nonmagnetic impurity\cite{yu}. Thus we come
to a conclusion that U-shaped DOS is due to the coexisting SDW and
SC order. We notice that the similar result has also been obtained
in the recent paper by Wan $et$ $al$ starting with $t-J$
model\cite{wang2}.

\begin{figure}[ht]
\epsfxsize=3.3in\centerline{\epsffile{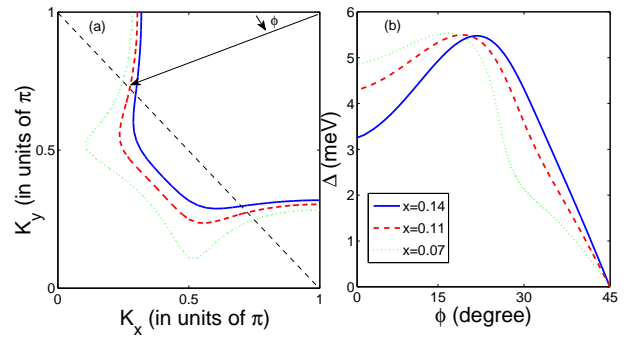}} \caption{(color
online) (a)The doping (see inset of (b)) dependence of the FS in the one
quarter of Brillouin zone. MBZ is denoted by the dashed line and $\phi$ is the FS angle. (b) corresponding non-monotonic gap as a function of the FS
 angle $\phi$. }
\end{figure}

\subsection{$d_{x^{2}-y^{2}}$-wave with a higher harmonic}

In this scenario, we only take the Hamiltonian $H_{SC}$ without AF SDW order
term. Based on a scenario in which superconductivity arises from the
exchange of AF spin fluctuations, it was argued that the maximum SC
gap is achieved near those momenta on the FS (hot spots) that are
connected by $Q=(\pi,\pi)$\cite{manske,blumberg,krotkov}. In
the electron-doped cuprates, the hot spots are located much closer
to the zone diagonal, leading to a non-monotonic behavior of the SC
gap. A good fit of $\Delta_{\bf k}$ to the experimental data is
achieved via the inclusion of a higher harmonic, such that
$\Delta_{\bf
k}=\Delta_{0}(\cos(k_{x})-\cos(k_{y}))/2-\Delta_{1}(\cos(3k_{x})-\cos(3k_{y}))/2$
with $\Delta_{0}=5.44$ meV and $\Delta_{1}=2.34$ meV ensures that
the maximum of $|\Delta_{\bf k}|$ along the FS is located at the hot
spots, as shown by the
red dashed curve in Fig.4(a) and Fig.4(b). By tuning the chemical
potential and $\Delta_{0}$, the doping dependence of FS and
correspondingly non-monotonic gap as a function of the FS angle
$\phi$ have also been shown in Fig.4. The other formulas on DOS and
LDOS remain.

\begin{figure}[ht]
\epsfxsize=3.5in\centerline{\epsffile{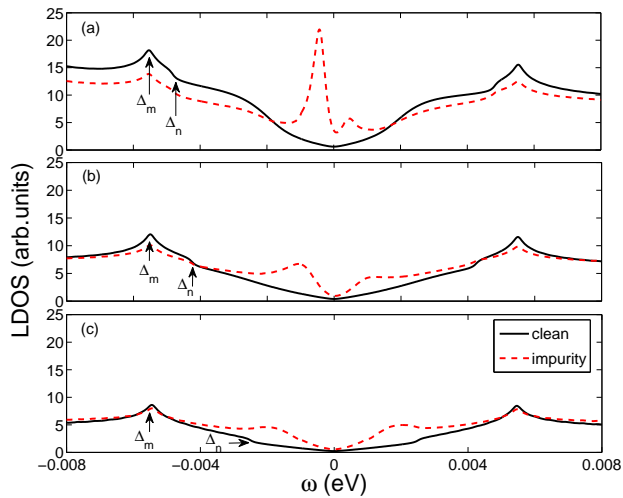}} \caption{(color
online) The LDOS for various doping in the clean case and the
nearest neighbor of a non-magnetic impurity site. (a)$x=0.14$,
(b)$x=0.11$, and (c)$x=0.07$. $\Delta_{n}$ and
$\Delta_{m}$ denote antinodal gap and maximum gap at the hot spot
respectively.}
\end{figure}

The doping dependence of LDOS has been plotted in Fig.5, where the
black solid curve denotes the clean case which is equivalent
to DOS, and red dashed curve denotes the nearest neighbor of
the impurity site. It has been shown that for all calculated doping, (i) the DOS at different
doping shows two Van Hove singularities at corresponding antinodal gap ($\Delta_{n}$) and
maximum gap at the hot spot ($\Delta_{m}$), reflecting the presence of a non-monotonic
$d_{x^{2}-y^{2}}$-wave gap, (ii) at low energies the DOS spectrum also
remains $d_{x^{2}-y^{2}}$-wave like and two
impurity-induced resonance states in LDOS locate symmetrically at positive
and negative energies and shift towards the low energy with the decreasing
doping. These features are similar to that shown in
the coexisting AF SDW and SC order scenario for the some small values $M$
(Fig.3(b) and Fig.3(c)). However, we notice that the U-shaped DOS which has
been clearly seen in the coexisting SDW and SC phase scenario at large SDW $M$
can not be obtained in the scenario based on
$d_{x^{2}-y^{2}}$-wave with a higher harmonic even at underdoped regime
(Fig.5(c)). Our results based on the $d_{x^{2}-y^{2}}$-wave with a higher
harmonic scenario are quite agreement with the recent doping dependence
of tunneling spectrum measurement\cite{dagan}, where a
non-monotonic gap order parameter for the whole doping range
studied has been shown. Therefore, the non-monotonic gap arising from the
$d_{x^{2}-y^{2}}$-wave with a higher harmonic due to the peculiar form of the pairing interaction, seems to be more reasonable than
the physical origin based on coexisting SDW and SC phase.

\section{Conclusion}

In summary, motivated by the probably physical origins of the
non-monotonic $d_{x^{2}-y^{2}}$-wave gap, we analyze the DOS and a nonmagnetic impurity effect in the electron-doped
cuprates superconductor assuming two different scenarios: the
$d_{x^{2}-y^{2}}$-wave superconductivity coexisting with AF SDW order versus
$d_{x^{2}-y^{2}}$-wave superconductivity with a higher harmonic
originating from the peculiar form of the pairing interaction. We
find that in both cases there exist two impurity-induced resonance
states at low energies. We also find that for the intermediate value
of the SDW gap, the DOS looks similar to that of the
$d_{x^{2}-y^{2}}$-wave gap with a higher harmonic. However, if the SDW gap is sufficiently large the
DOS looks more conventional s-wave like instead of the
non-monotonic $d_{x^{2}-y^{2}}$-wave gap behavior. This obvious
difference from that of the $d_{x^{2}-y^{2}}$-wave gap with a higher
harmonic even in the underdoped regime, will help to differentiate
the two above scenarios and make sure the physical origin of the
unusual non-monotonic gap in the electron-doped cuprates. The recent
doping dependence of tunneling spectrum measurement\cite{dagan}
provides the strong evidence of the non-monotonic gap arising from
the $d_{x^{2}-y^{2}}$-wave with a higher harmonic, and is qualitatively agreement with our theoretical
results. However, the earlier point contact tunneling spectrum and
penetration depth measurement show the U-shaped DOS, suggesting the
coexisting SDW and SC order. Therefore, in order to reach the
consensus on this topic, our results strongly suggest the further
doping dependence of tunneling spectrum measurement, especially in
the underdoped regime, is needed to be carried out as soon as
possible.

\section*{Acknowledgments}

We thank Ilya Eremin and Jun Chang for many useful discussions. Y.L. acknowledges support from National Natural Science Foundation
of China under Grand Nos.10404001.

\end{document}